\documentclass{fundam}

\publyear{24}
\papernumber{2102}
\volume{185}
\issue{1}

\usepackage{url} % takes care of hyperlinks, preferred over hyperref
\usepackage[ruled,lined]{algorithm2e}% provides Algorithm environment
\usepackage{graphicx}% allows for inclusion of graphic files (figures)
\usepackage{amsmath}% for \mathbb{Z}
\usepackage{amssymb}% for \mathbb{Z}
\usepackage[lofdepth,lotdepth]{subfig}
\usepackage{caption} 
\usepackage{listings}
\usepackage{tikz}
\usetikzlibrary{automata, positioning, arrows}

\begin{document}

\title{Mechanical Self-replication}

\address{ralph.lano@th-nuernberg.de}

\author{Ralph P. Lano\\
Technische Hochschule Nürnberg - Georg Simon Ohm\\ Keßlerplatz 12, 90489 Nürnberg, Germany\\
ralph.lano@th-nuernberg.de}

\maketitle

\runninghead{Ralph P. Lano}{Mechanical Self-replication}

\begin{abstract}
This study presents a theoretical model for a self-replicating mechanical system inspired by biological processes within living cells and supported by computer simulations. The model decomposes self-replication into core components, each of which is executed by a single machine constructed from a set of basic block types. Key functionalities such as sorting, copying, and building, are demonstrated. The model provides valuable insights into the constraints of self-replicating systems. The discussion also addresses the spatial and timing behavior of the system, as well as its efficiency and complexity. This work provides a foundational framework for future studies on self-replicating mechanisms and their information-processing applications.
\end{abstract}

\section{Introduction}

One of the first models proposed for self-replication was the model of von Neumann [1].  It primarily consists of a constructor automaton (A), copier automaton (B), and control automaton (C), together with their respective descriptions $\Phi$.  Some time later, these automata were identified in living organisms.  Here, the constructor automaton (A) can be associated with ribosomes, which perform protein synthesis in the process of translation, and the copier automaton (B) can be associated with RNA polymerase, which makes copies of the descriptions $\Phi$ in a process called transcription.  The descriptions $\Phi$ can be associated with the genetic code, RNA, or DNA [2-5].

Mechanical self-replicating and self-reproducing structures have been extensively presented and discussed in the literature in various levels of detail [6-8].  More practical implementations were suggested in [9-12], for instance.

The goal of this study is to develop a simple mechanical system capable of self-replication.  The system shown in Fig.1 consists of six types of interconnected small machines, each of which performs one task.  The machines themselves were built from five types of basic building blocks.

\begin{figure}[htbp]
\centering
\includegraphics[width=\linewidth]{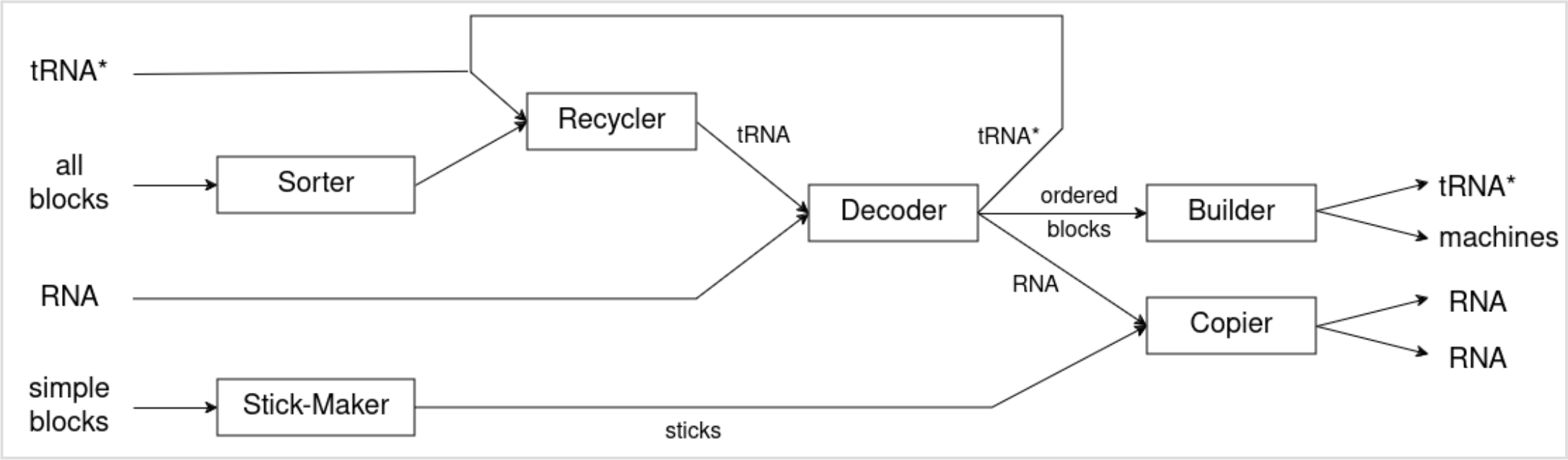}
\caption{Overview}
\label{fig:Fig_01}
\end{figure}

On the left side are the inputs to the system: the basic building blocks (simple blocks and all blocks), the building instructions (RNA), and something that we call tRNA*, a simple structure made out of basic building blocks.  On the right side, we see the outputs, which are copies of the machines, copies of the tRNA*, and a copy of the RNA.

The copying mechanism consists of the Stick-Maker and the Copier.  The Stick-Maker takes simple blocks and makes sticks out of them.  The Copier then uses these sticks to generate copies of RNA.  The construction mechanism consists of the following machines: the Sorter, the Recyclers, the Decoder, and the Builder.  The Sorter distinguishes the different block types, and forwards the sorted blocks to the Recyclers.  For each block type, there is one Recycler that matches a particular block type with a specific tRNA* to produce a filled tRNA.  The filled tRNA are then forwarded to the Decoder.  The Decoder uses these tRNA, matches them against the RNA, and produces a stream of ordered blocks.  The Decoder does not modify the RNA and leftover tRNA* is returned to the Recyclers.  The stream of ordered blocks is then used by the Builder to build machines, structures, and tRNA*. 

In von Neumann's language, the Stick-Maker together with the Copier corresponds to automaton B, whereas the combination of the Sorter, Recyclers, Decoder, and Builder corresponds to automaton A.

\subsection{Block Types}

Our mechanism has five building block types: temporary blocks, building blocks, active blocks for movement and gluing, and blocks required for sorting.  Figure 2 shows the different block types.  There is essentially an unlimited supply of them.  From these blocks, all the machines described above are built, as well as the instruction code to build these machines, the RNA.  In their inactive state, all have the same shape and size.  All blocks can be glued to other blocks to form larger structures, which are referred to as compounds.

\begin{figure}[htbp]
\centering
\includegraphics[scale=0.175]{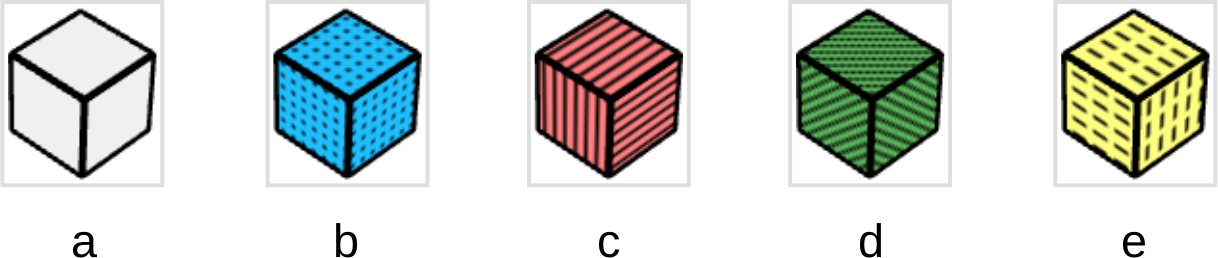}
\caption{Block types}
\label{fig:Fig_02}
\end{figure}

No assumptions are made regarding the nature or size of these blocks, except that they are relatively small.  They can be biological, chemical, or mechanical in nature.  The expansion of the Mover can be caused biologically, chemically, mechanically, through magnetic or electrostatic repulsion, or hydraulically.  The glue mechanism could be a form of actual glue, a type of chemical bond, or mechanically similar to a barb or Velcro.  This could be more sophisticated, such as the Penrose mechanism of interlocks [9].  It could involve magnetic or electrostatic attraction or repulsion, an interlock mechanism, as found in proteins, or click chemistry.

Dissolvable: The Dissolvable block (Fig.2(a)) is a temporary block type required only during construction, and it dissolves a relatively short time after construction.  It is used as a filler material during construction, similar to the dissolvable support structures in 3D printing.

Simple: The Simple block (Fig.2(b)) is the basic building material.  It is inert, does not change shape or size, and has the ability to be glued to other blocks as its main property.  Gluing is initiated from the outside by the Gluer block.

Mover: The details of the Mover block (Fig.2(c)) are shown in Fig.3.  Mover blocks can be in a normal state (Fig.3(a)) or an expanded state (twice their original size), as shown in Fig.3(b).  They expand in a fixed predetermined direction.  By expanding, the Mover can move other blocks and compounds.  The direction of the Mover defines which blocks are moved, as shown in Figs.3(c) through (e).

\begin{figure}[htbp]
\centering
\includegraphics[scale=0.175]{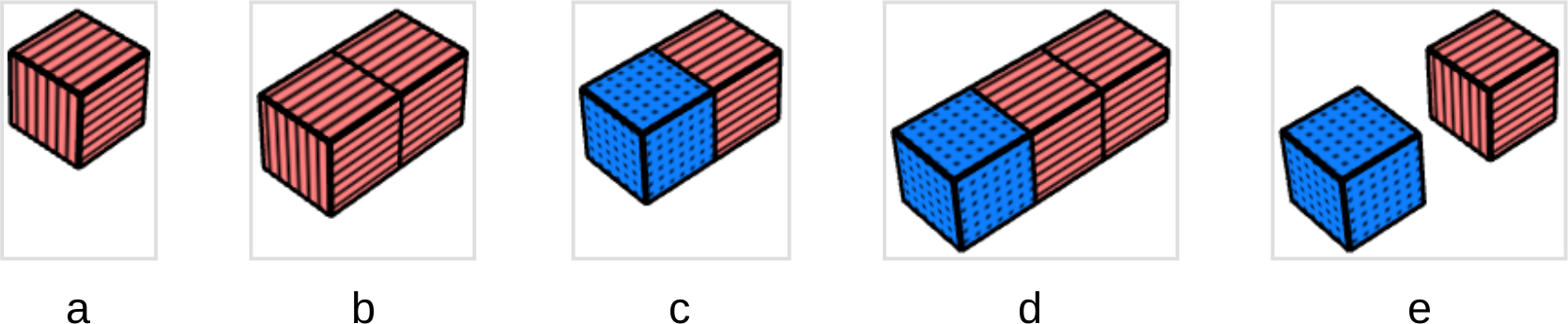}
\caption{Mover block}
\label{fig:Fig_03}
\end{figure}

The timing mechanism of the expansion is important for understanding.  The expansion of the Mover blocks within one machine is synchronized relative to each other.  There is no time synchronization between the different machines.  A given Mover will expand at its tick-time ($tt$) and stay expanded for an amount of time ($et$), after which it will contract again.  Both the tick- and expansion times can be chosen as integers.  For the mechanism presented here, $tt$ has a value between zero and nine, and $et$ has a fixed value of two.

A Mover block can move a significant number of other blocks, but a critical assumption is that Mover blocks cannot break compounds apart.  During construction and transport, Mover blocks are in an inactive state, meaning they do not expand.  This can easily be accomplished if the energy source required for expansion is not available during transport and construction.  Because there are six different directions, and we allow for ten different tick-times, there are at most 60 different Mover block types.  However, a significantly smaller number is required to construct the machines presented here.

Gluer: Details regarding the Gluer blocks (Fig.2(d)) are shown in Fig.4.  During construction and transport, Gluer blocks are in an inactive state.  Each Gluer block has a fixed, predetermined direction $d$, that defines its region of influence.  Hence, in principle, there are six Gluer blocks.  Assuming this direction to be the positive z-direction (upward), Fig.4(a) depicts an example of a 9-neighborhood (Moore) region of influence.  Other neighborhoods are also possible, but all machines presented in the following use Gluer blocks with a 9-neighborhood region of influence.

\begin{figure}[htbp]
\centering
\includegraphics[scale=0.175]{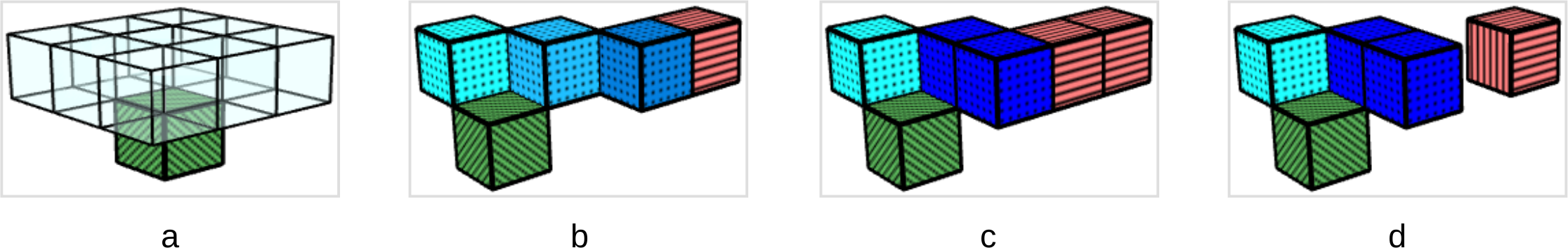}
\caption{Gluer block}
\label{fig:Fig_04}
\end{figure}

To briefly demonstrate the action of the Gluer, consider Figs.4(b)-(d).  The Gluer's region of influence in this figure is upward, having a 9-neighborhood.  Once the Mover expands, all the shown Simple blocks are in the region of influence.  Because the Gluer only glues blocks that share a side (no diagonals), only two of the blocks depicted (dark blue) were glued together.  Note, that the Mover never enters the volume of influence.

If a block enters the region of influence of a Gluer block, and if another block is close (that is, shares a common side) and in the region of influence, these two blocks are glued together.  Several blocks that are glued together are called a compound.  Compounds move as a whole, meaning that the positions of the blocks within a compound do not change.  In addition, a compound cannot break apart.  Compounds that contain active blocks, such as the Mover or the Gluer, are called machines.  Gluing may require energy or may be catalytic.  The Gluer block itself can be glued to other blocks by another Gluer, except for the active side of the Gluer block.  Nothing can be glued to the active side of a Gluer block.

Sorter: The Sorter blocks (Fig.2(e)) are used to distinguish between the different block types.  Figure 5 shows a detailed view of a possible choice for Sorter blocks, where each block has a ridge extruding at a different position.  Naturally, these ridges must not be extruded during transport and construction.

\begin{figure}[htbp]
\centering
\includegraphics[scale=0.175]{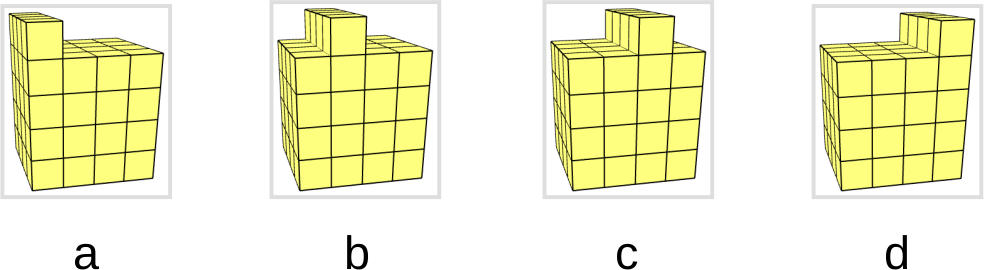}
\caption{Sorter blocks}
\label{fig:Fig_05}
\end{figure}

For the Sorter blocks to work, we must require that every block type has a unique marking through grooves at the bottom, as indicated in Fig.6, also a detailed view.  In this way, we can use the Sorter blocks to distinguish the different block types.

\begin{figure}[htbp]
\centering
\includegraphics[scale=0.175]{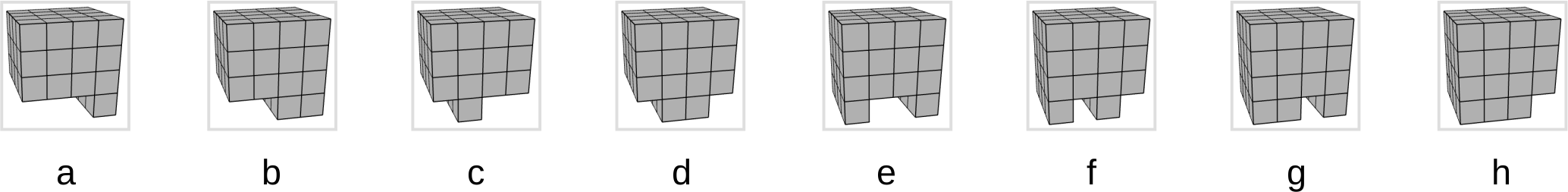}
\caption{Markings}
\label{fig:Fig_06}
\end{figure}

More details of the scheme used will be explained later, when we give the details of the Sorter machine.

\subsection{Codons, RNA, and tRNA}

The first issue that needs to be solved is the representation of information.  Inspired by biology, we introduce the concept of a basic information unit, the codon.  Figure 7 depicts 2-codons.  The "2" stands for two blocks width or two bits of information.  Looking at the upper extrusions, one can recognize the binary numbers 00 in Fig.7(a), 01 in Fig.7(b), 10 in Fig.7(c), and 11 in Fig.7(d).  Clearly, this can be extended to n-codons, where n is an integer larger than 2.

\begin{figure}[htbp]
\centering
\includegraphics[scale=0.175]{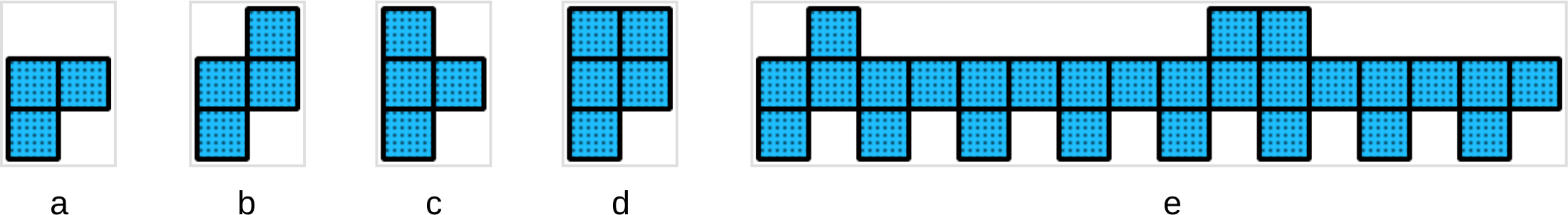}
\caption{Codons and RNA}
\label{fig:Fig_07}
\end{figure}

Combining codons by gluing them allows us to store information in the RNA tape depicted in Fig.7(e).  Conventionally, the tape is read from left to right.  The markings in the lower part of the tape allow us to distinguish the direction in which the tape is to be read, assuming n to be even.  And they allow for easy transport.

Reading information from the tape is performed using tRNA.  Again, this idea is inspired by biology.  We distinguish between tRNA* and tRNA: the former is empty, whereas the latter is filled with building material.  Figs.8(a)-(d) depict 2-tRNA*, whereas Figs.8(e)-(h) depict 2-tRNA.  At the bottom, it matches the corresponding codon, and at the top, there is an empty space left to carry a block.  For instance, one could choose that 2-tRNA* in Fig.8(a) carries a Dissolvable block, which is shown in Fig.8(e), etc..  For every codon, there is a corresponding tRNA*.  For instance, 2-tRNA* in Fig.8(a) matches the 2-codon shown in Fig.7(a), etc..  This particular association of block type with tRNA type is arbitrary, but should be fixed.  This can be extended to n-tRNA, where n is an integer larger than two, thus allowing for distinguishing between more different block types.  The notch in the middle of the tRNAs is required for transport.

\begin{figure}[htbp]
\centering
\includegraphics[scale=0.175]{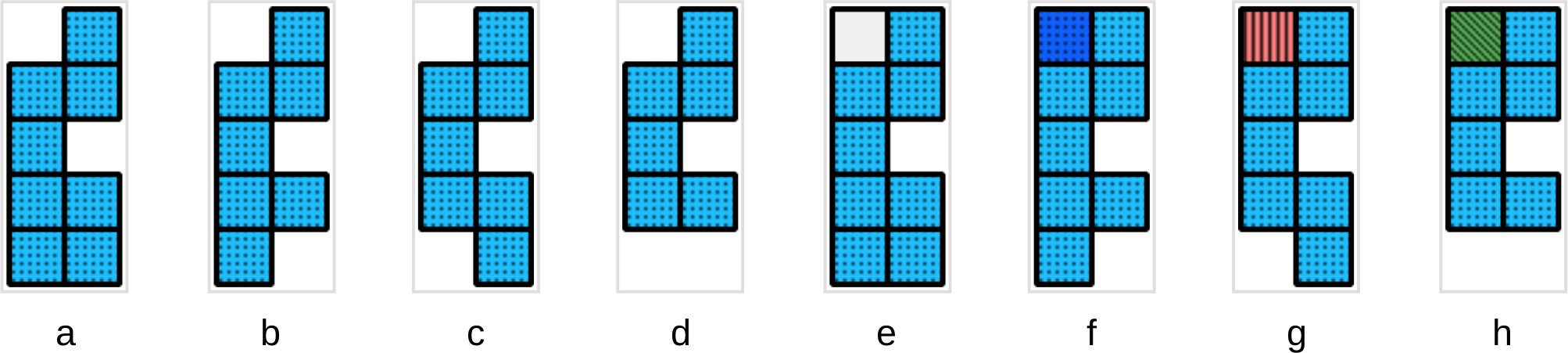}
\caption{tRNA* and tRNA}
\label{fig:Fig_08}
\end{figure}

If we extend the encoding scheme to n-codons, we must do the same for tRNA.  The minimum value for n can be estimated by counting the different block types needed to construct the machines presented later.  We need to distinguish the different block types: there is one Dissolvable, one Simple, and at least four Sorter block types.  Assuming that we can manage with two Gluer types and eight Mover types, n would be equal to four.  If we need to distinguish more different block types, n would need to be larger, and the number of Sorter blocks would also need to be larger.

\subsection{Machine Description Language (MDL)}

To describe structures such as the tRNA above or the machines presented later, we need to introduce a machine description language (MDL).  The MDL uniquely describes a machine in terms of its blocks.  To represent a particular block, we need to describe its type: we use 'b' for Simple block, 'd' for Dissolvable block, 's' for Sorter block, 'G' for Gluer block, and 'M' for Mover block.  The Gluer blocks and Mover blocks have a direction that is represented by a number between zero and five.  In addition, the Mover blocks have timing information, which is represented by a number between zero and nine.  Hence, "M28" represents a Mover block ('M') that points in the positive z-direction ('2') and expands at tick time 8, or "G2\_" represents a Gluer block pointing in the positive z-direction.  To distinguish between the different Sorter types, we use the letter 's' followed by a number, indicating which Sorter it is; for instance, "s2\_" refers to Sorter number 2 (Fig.5(b)).  In using uppercase letters, we indicate active block types. Finally, we used "\_\_\_" to denote an empty space.

As an example, consider the MDL (Fig.9) that describes the Builder machine defined later in this paper.  We sliced the machine in the z-direction to obtain x-y planes.  A line denotes the x-direction; with a new line, we indicate the y-direction.  In the given example, the first line consists of seven simple blocks.  The second line goes one down in the y-direction and consists of six simple blocks, with a Mover block pointing in the upward z-direction, expanding at tick-time 8.  The end of the plane is denoted by an empty line.  Comments are ignored. Therefore, we describe machines one layer at a time.

\begin{figure}[htbp]
\centering
\includegraphics[scale=0.175]{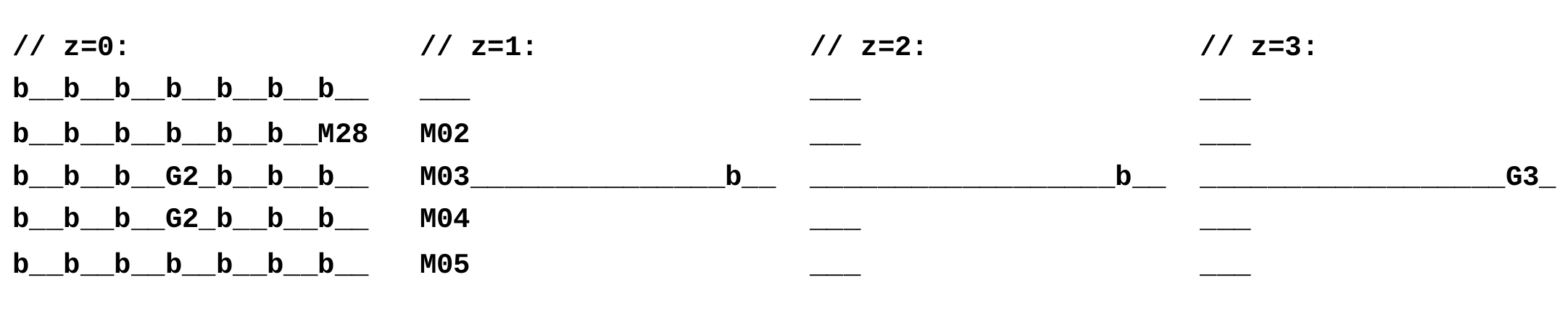}
\caption{MDL of Builder}
\label{fig:Fig_09}
\end{figure}

It is important to note that for every compound or machine, we require one separate MDL file.  We do not describe two or more compounds or machines in one MDL file.

\section{Method}

We must make certain assumptions about the environment in which our blocks and machines exist and the rules they follow.  These assumptions are inspired by the interior of living cells, which consists of a fluid-like environment, allowing its components to float freely and unhindered by gravity.  Hence, we neglected both gravity and viscosity.  An essential feature is that these blocks can be glued together to form compounds.  These compounds are very stable and do not break, similar to the strong chemical bonds.  Additionally, we may need a weak bond, similar to a van der Waals-type bond, for tRNA to carry their load.  It should be noted that there is no rotation in the described mechanism.

These assumptions were implemented in a computer program, and all the machines presented in this paper were simulated.  In the following section, we describe the set of rules implemented in the program.  Most of these rules are self-evident; however, some are specific to this particular simulation.

\subsection{Rules}

\begin{enumerate}
\item Environment: Our world is three-dimensional with coordinates $x$, $y$, and $z$, where $x, y, z \in \mathbb{Z}$. In other words, only integer coordinates are allowed.  In addition, there is a global time $t$, where $t$ is a non-negative integer that starts at zero and is incremented by one during the simulation.

\item Blocks: The basic building units are blocks. There are five types of blocks: Dissolvable, Simple, Sorter, Gluer, and Mover.  All blocks have the size of a unit cube, that is, width = height = depth = 1.  The exception is the Mover block, which can expand by one unit in a predefined direction.  No two blocks can be placed in the same position.

\item Block Movement: Blocks can be moved by one unit in either the positive or negative $x$, $y$, or $z$ direction.  Diagonal movements are not allowed.  No rotation is allowed.

\item Compound: All blocks are organized in compounds.  A compound can consist of one or more blocks.  Every block must be part of one and only one compound. 

\item Compound Modification: Only one block can be added to a compound at a time.  This modification can only occur via the Gluer block.

\item Compound Movement: Compounds move as a whole, meaning that the positions of the blocks within a compound cannot be changed.

\item Mover: A Mover block has two additional properties: tick-time ($tt$) and direction ($d$). Tick-time is a number between zero and nine. The direction is a number between zero and five, where zero indicates the positive x-direction, one indicates the positive y-direction, two indicates the positive z-direction, three indicates the negative x-direction, four indicates the negative y-direction, and five indicates the negative z-direction. This defines the active side of the Mover.

\item Mover Expansion: The Mover will expand by one unit in the direction $d$ at tick-time $tt$.  To be more precise, the expansion occurs when the global time $t$ modulo the tick-time $tt$ is zero.  After two ticks, the Mover contracts again to its original size.  Hence, the expansion occurs periodically every ten ticks, starting when $t \mod tt == 0$, and lasts for two time units.  If another block is on the active side of the Mover block before expansion, it will be moved during expansion in the direction of $d$ by one unit, except in the case of rule 9.

\item Self-Movement: A Mover cannot move itself.  This means that during expansion, the non-expanding part of the Mover cannot change its position.  If that were to happen, the Mover does not expand.

\item Mover Strength: The Mover can move a large number of other blocks through its expansion.  However, compounds cannot be broken apart.  If the expansion of a Mover was to cause a compound to break apart, the Mover would not expand.

\item Gluer: A Gluer block has an additional property of direction ($d$).  It is defined in the same way as for the Mover block and defines the active side of the Gluer block.  The volume next to it is called the volume of influence.

\item Gluing: If two blocks are in the region of influence of the Gluer block, are not part of the same compound, and the distance between their centers is one (that is, they share a side), then the two compounds to which these two blocks belong are glued together.

\item No Gluing to Active Side: Nothing can be glued to the active side of a Gluer block.

\item Transport: During transport and before final assembly, the Mover, Gluer, and Sorter blocks are in an inactive state, meaning that the Mover block does not expand, the Gluer block does not glue, and the Sorter block is not extruded yet.
\end{enumerate}

Although inspired by the physical world, our model makes some significant simplifications; for example, there are no rotational degrees of freedom.

\section{Patterns}

To understand how our mechanism and machines work, we first consider simple examples.  These are reusable patterns that appear in several machines.

\subsection{Conveyor}

Figure 10 depicts the workings of the move-by-two Conveyor machine.  The move-by-two Conveyor consists of one Mover block that expands in a perpendicular direction and two Mover blocks that expand in the direction defined by the machine.  Figure 10(a) shows a machine in its resting state.  First, the perpendicular Mover block expands (Fig.10(b)), after which the other two Mover blocks expand (Fig.10(c)).  Then the perpendicular Mover block contracts (Fig.10(d)), and finally, the other two Mover blocks contract, thus returning the Conveyor to its initial state (Fig.10(a)).

\begin{figure}[htbp]
\centering
\includegraphics[scale=0.175]{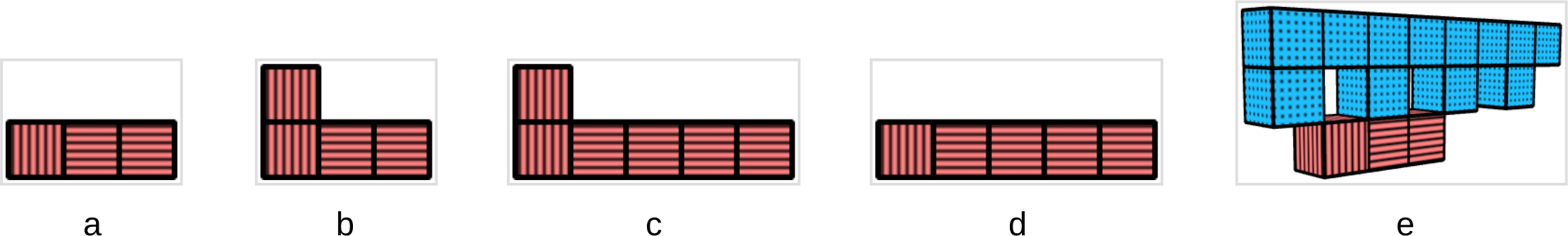}
\caption{Conveyor}
\label{fig:Fig_10}
\end{figure}

Figure 10(e) depicts a move-by-two Conveyor (red) together with a Belt (blue).  If one considers the move-by-two Conveyor to be fixed, then the Belt is being moved by the Conveyor.  The Belt is reminiscent of actin, a family of proteins that form microfilaments in the cytoskeleton of eukaryotic cells.  The Conveyor in this context is reminiscent of myosins, a family of motor proteins used in muscle contraction and in a wide range of other motility processes in eukaryotes.

\subsection{Walker}

Figure 11 depicts the workings of a move-by-two Walker.  The move-by-two Walker consists of two Mover blocks expanding perpendicularly and two Mover blocks expanding in the direction defined by the machine (longitudinally).  In addition, a Simple block (blue) was used as a filler with no functionality.  Figure 11(a) shows a machine in its resting state.  First, the first perpendicular Mover block expands (Fig.11(b)), after which the two longitudinal Mover blocks expand (Fig.11(c)).  Next, the first perpendicular Mover block contracts, while the last perpendicular Mover block expands (Fig.11(d)).  Then, the two longitudinal Mover blocks contract (Fig.11(e)), and finally, the last perpendicular Mover block contracts, returning to the initial state (Fig.11(a)).

\begin{figure}[htbp]
\centering
\includegraphics[scale=0.175]{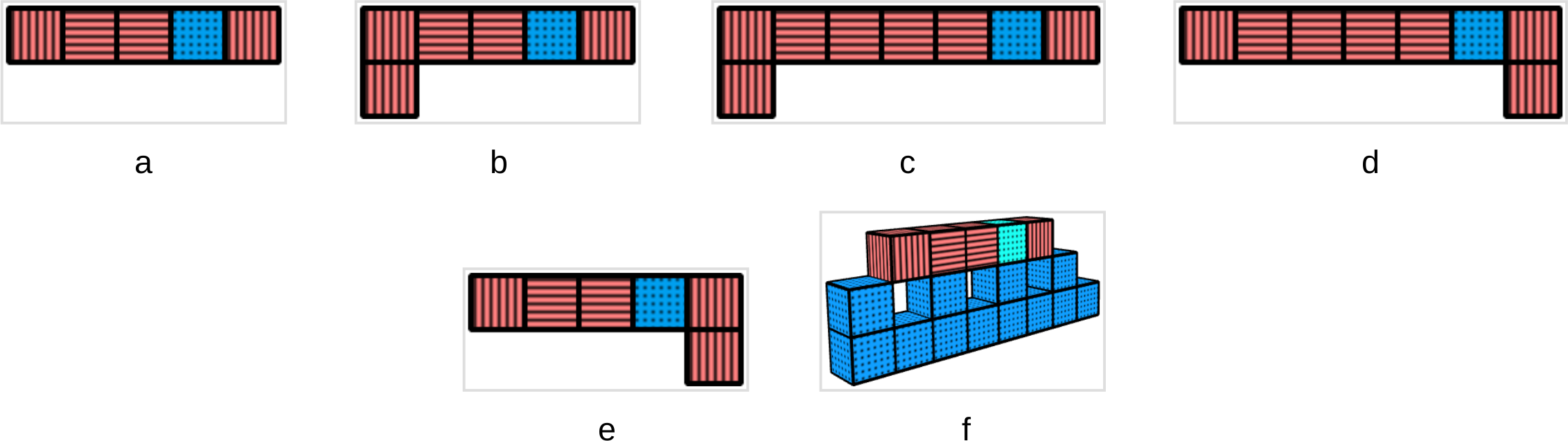}
\caption{Walker}
\label{fig:Fig_11}
\end{figure}

Figure 11(f) depicts a move-by-two Walker (red) and a Track (blue).  If one considers the Track to be fixed, then the move-by-two Walker machine moves along the Track.  This is reminiscent of the kinesin protein, a motor protein found in eukaryotic cells.

\subsection{Redirect}

Redirection can be achieved by using a simple Mover.  In the example shown in Fig.12, a move-by-two Conveyor (red) moves a tape of 2-tRNA (transparent) from right to left.  One extra Mover pushing from the back was used to create a constant stream of blocks.

\begin{figure}[htbp]
\centering
\includegraphics[scale=0.175]{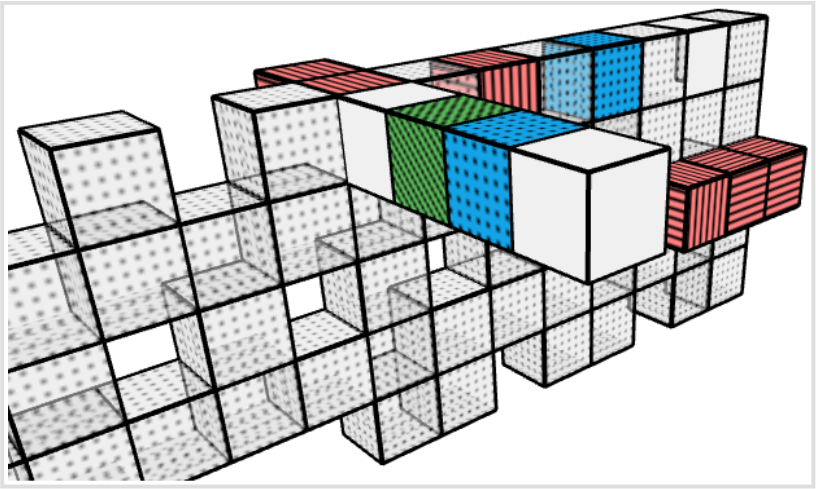}
\caption{Redirect}
\label{fig:Fig_12}
\end{figure}

\subsection{1-to-2 Converter}

A 1-to-2 Converter, shown in Fig.13, converts a stream of blocks of width one into a stream of blocks of width two.  This demonstrates the use of a Blocker, as shown in magenta.  We have a stream of blocks coming from the right side.  The Converter consists of three Mover blocks connected through a frame.  The upper Mover block will try to push down through the Lever (dark blue), but is prevented from doing so by the Blocker.  Only when the Blocker is pushed back through the incoming stream of blocks will the upper Mover block be able to push down two blocks.  Once that is done, a Mover in the back (not shown) will push the Lever up again, and another Mover, shown to the left, will push the Blocker back to its original position.  It is clear that this can be easily extended to any width.

\begin{figure}[htbp]
\centering
\includegraphics[scale=0.175]{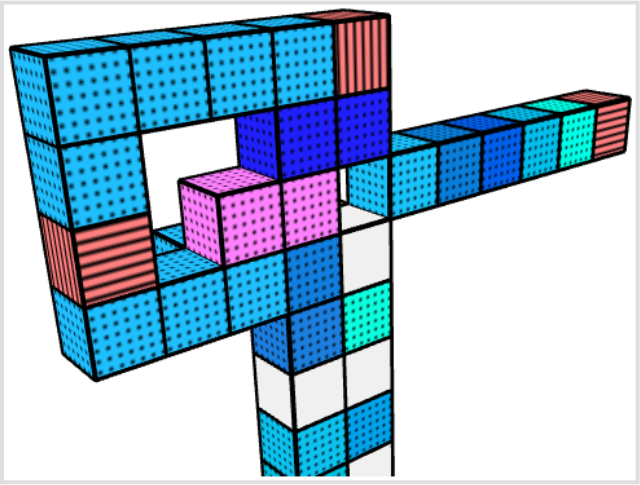}
\caption{1-to-2 Converter}
\label{fig:Fig_13}
\end{figure}

This is our first example of a Blocker pattern: only after a certain condition is met will something occur.  In the above case, at least two blocks must be present before they are pushed down.  On the one hand, this is a form of a counter, and on the other, a form of a conditional.

\subsection{Matcher}

A Matcher is a central component of several machines.  Let us first examine the Matcher by itself, as shown in Figs.14(a) and 14(b).  It consists of a frame (light blue) and a Mover (red).  If we place a codon (dark blue) in the lower front part and an anti-codon (magenta) in the upper back part, two things can happen: codon and anti-codon match (Figs.14(c)-(d)) or codon and anti-codon do not match (Fig.14(e)).

\begin{figure}[htbp]
\centering
\includegraphics[scale=0.175]{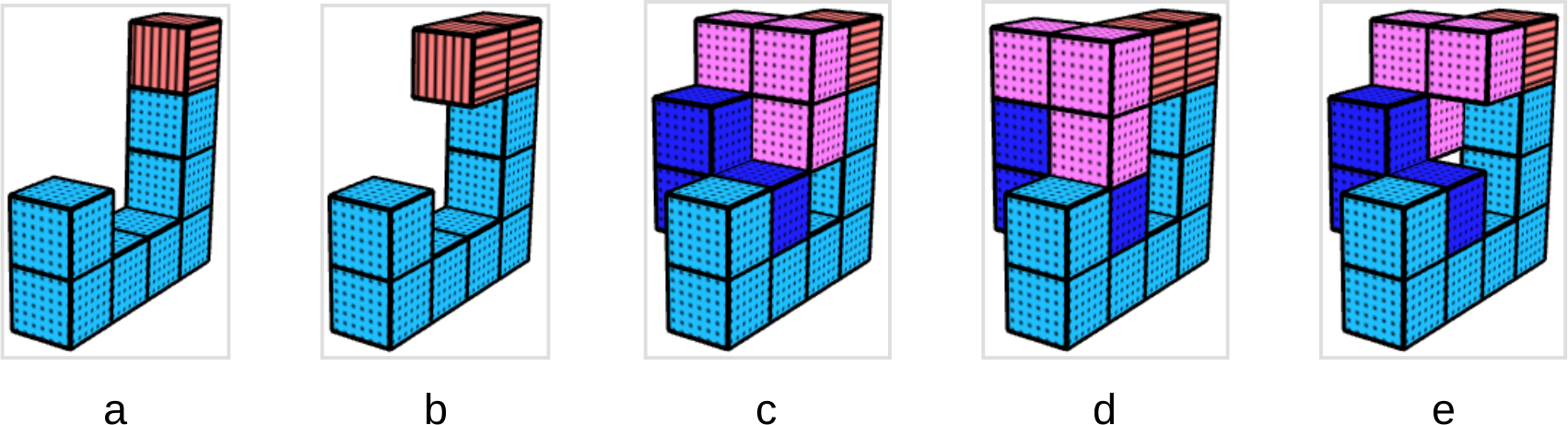}
\caption{Matcher}
\label{fig:Fig_14}
\end{figure}

If they do not match, the anti-codon will remain where it was.  If we combine this with a move-by-two Conveyor that moves anti-codons from left to right, we see that this will move only the anti-codons that match the given codon.  This can be used to sort codons, to pick the correct tRNA, or to copy a given RNA tape.  Notice that the Mover is not strong enough to break the chemical bond of the frame.  This is a very critical assumption.

\subsection{Back-and-Forth}

Figure 15 depicts the Back-and-Forth machine, which we could also refer to as two-state machine.  It consists of four Movers arranged as shown and a middle piece shown in blue.  The Movers are connected through a frame (not shown) such that their relative positions are fixed.  First, the middle piece is moved to the left (Fig.15(b)).  Then, it is moved back (Fig.15(d)), then it is moved to the right (Fig.15(f)), and finally, it moves forward again (Fig.15(h)).  The two states are the right and left states or the forward and backward states.

\begin{figure}[htbp]
\centering
\includegraphics[scale=0.175]{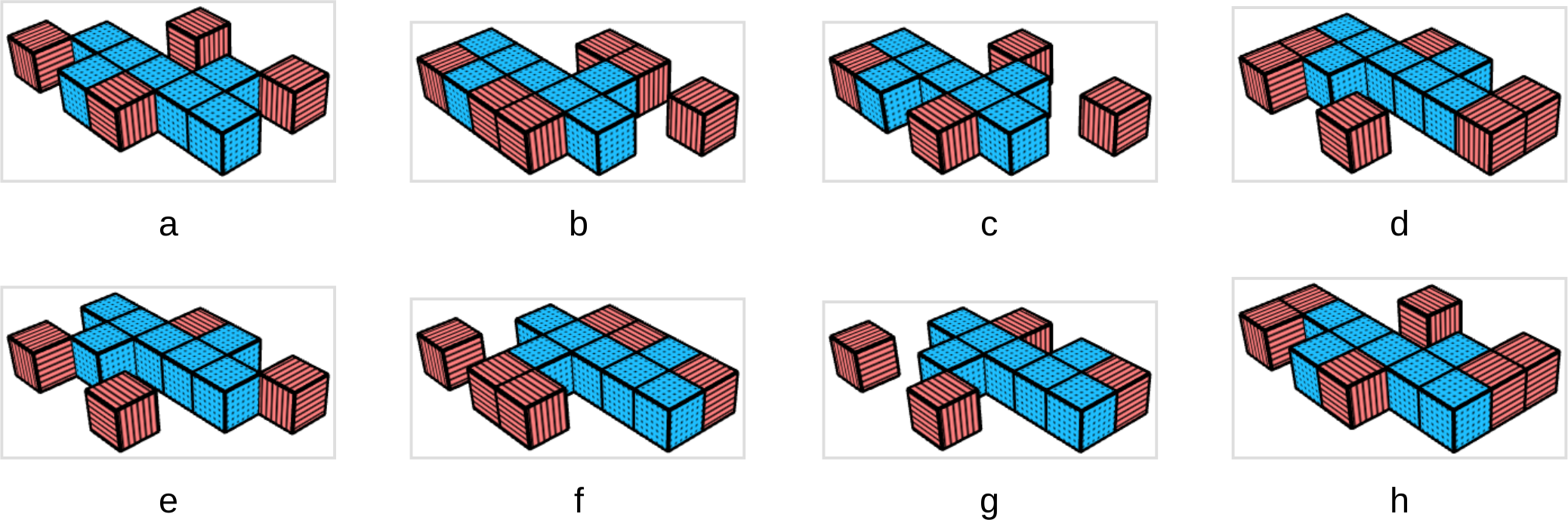}
\caption{Back-and-Forth}
\label{fig:Fig_15}
\end{figure}

It can also be used as a modulo-two counter or to cause delays.  In addition, it can be extended to a modulo-three counter or any number for that matter.  For the modulo-three counter, we would have two Movers in the front and one move-by-two Conveyor in the back, and the middle piece needs to be of a slightly different shape.

\section{Machines}

We are now ready to implement the machines shown in Fig.1.  We begin with the copy mechanism (automaton B) and then proceed to the build mechanism (automaton A).

\subsection{Stick-Maker}

For the Copier machine to work, we need to produce 3-Sticks, as shown in Fig.16(a).  3-Sticks are simply three Simple blocks glued together ("b\_\_b\_\_b\_\_"), which we can create with the Stick-Maker, as shown in Fig.16(b) through (e).  This process assumes that three Simple blocks and one Dissolvable block come from above or the side.  The move-by-four (red) then moves them forward over the Gluer block (green), which glues them together.  The Dissolvable blocks are essential because they allow sticks to separate after a while.

\begin{figure}[htbp]
\centering
\includegraphics[scale=0.175]{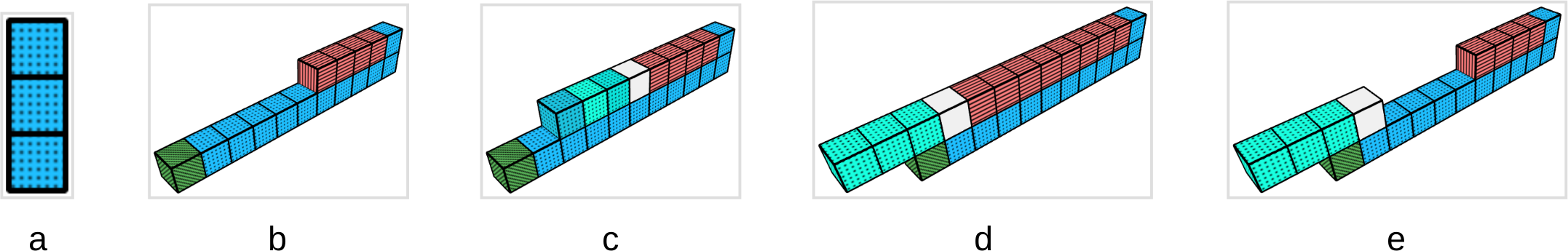}
\caption{Stick-Maker}
\label{fig:Fig_16}
\end{figure}

Figure 16(b) shows an empty Stick-Maker for processing Simple blocks into 3-Sticks.  The Stick-Maker has a Gluer (green) at one end and a move-by-four (red) at the other end, positioned one block higher relative to the Gluer.  Figure 16(c) shows the first step, where three Simple blocks and one Dissolvable block are fed to the Stick-Maker from above or the side.  For instance, a modified 1-to-2 Converter can achieve this.  In the second step (Fig.16(d)), the move-by-four expands and pushes the inserted blocks over the Gluer block, which glues them together to form 3-Sticks.  Finally, in Fig.16(e), the move-by-four contracts, returning the Stick-Maker to its original state.  Repeating this process several times creates a long chain of 3-Sticks temporarily connected to each other by Dissolvable blocks, which, as their name implies, dissolve after some time, leaving us with the desired 3-Sticks.  These can then be forwarded to the Copier.

Alternatively, one could make the sticks four blocks long and then use a simple Move-by-Two at the very end to move the created sticks out of the Gluer's region of influence, eliminating the need for the Dissolvable block.

\subsection{Copier}

For the Copier presented here, we modified the RNA tape slightly, as shown in Fig.17.  It is made up of 3-Sticks only, but still contains the same amount of information.

\begin{figure}[htbp]
\centering
\includegraphics[scale=0.175]{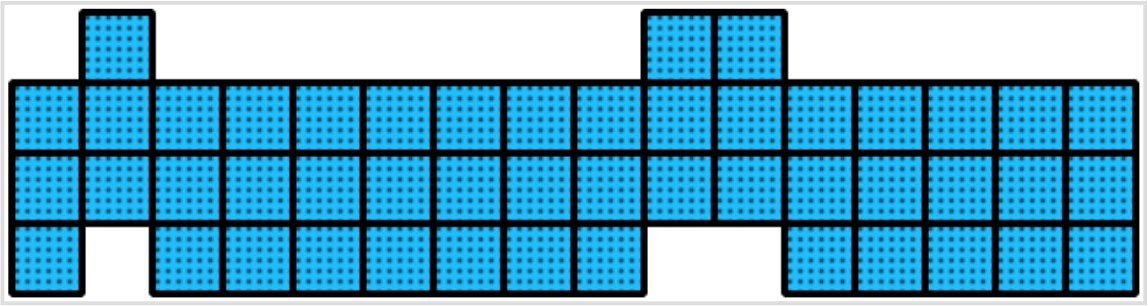}
\caption{Modified RNA Tape}
\label{fig:Fig_17}
\end{figure}

The Copier is depicted in Fig.18(a), consisting of a frame (blue) connecting a Gluer, two Movers, and one Move-by-Two.  RNA tape (transparent) comes from the right.  For this version of the Copier to work, we must assume that the start codon of the tape is in the form shown; otherwise, the Copier cannot move the tape forward.

\begin{figure}[htbp]
\centering
\includegraphics[scale=0.175]{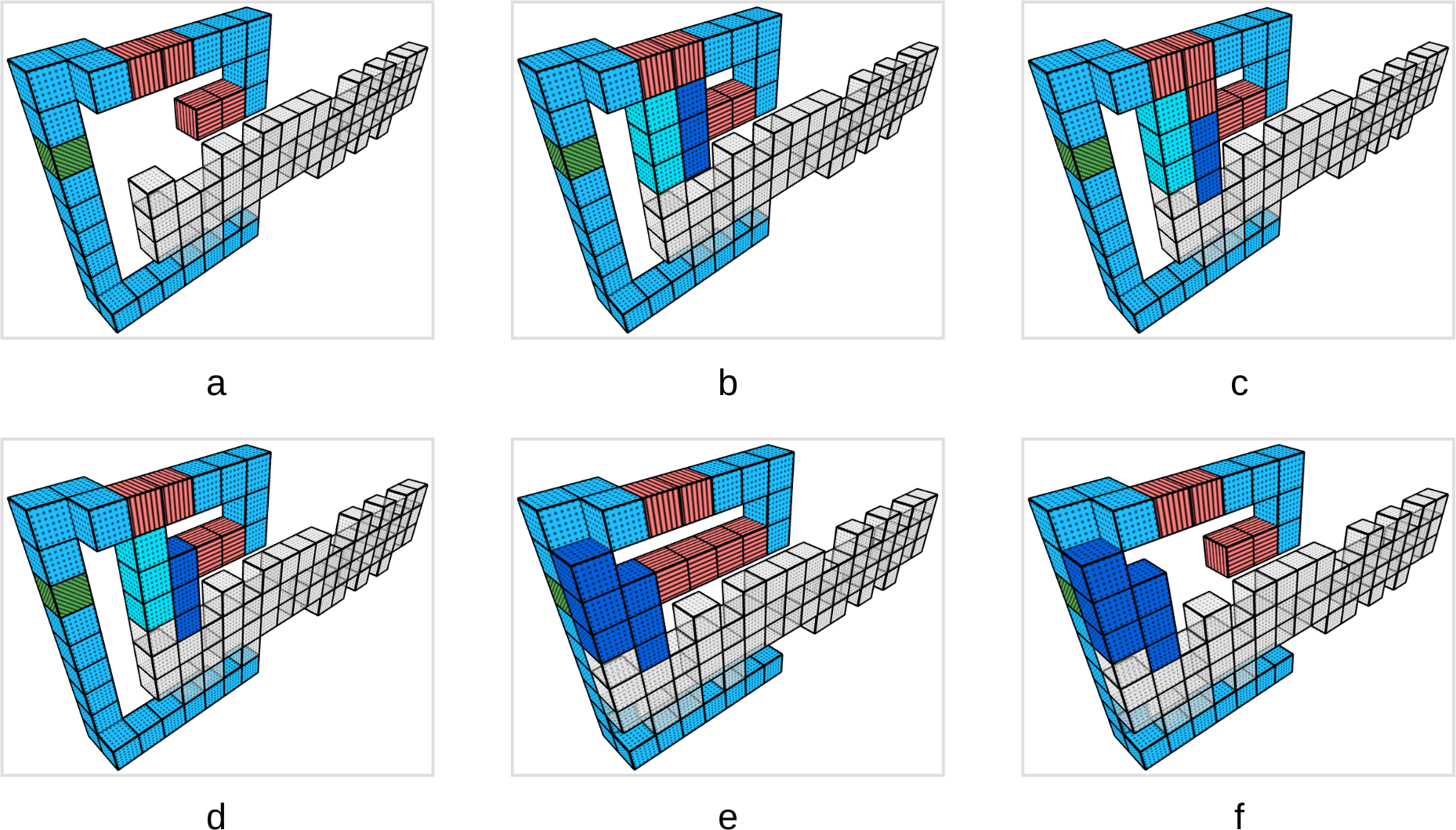}
\caption{Copier}
\label{fig:Fig_18}
\end{figure}

Figure 18(b) shows the Copier after two 3-Sticks were fed into it from behind, just under the two Movers and above the RNA tape.  In the next step (Fig.18(c)), the two Movers try to push the 3-Sticks down, matching them with the RNA tape.  In Fig.18(d), the two Movers have retracted.  Now, in Fig.18(e), the Move-by-Two can expand, pushing the two 3-Sticks to the left, and because of the shape of the start codon, the whole RNA tape moves with them.  More importantly, the 3-Sticks get pushed past the Gluer block, thereby being glued together.  Fig.18(f) shows the Copier after one cycle of operation, ready to pick up the next group of 3-Sticks.

Several types of Copier machines are possible.  The advantage of the one presented here is that, in one pass, it can produce a positive copy of the provided RNA tape.

\subsection{Builder}

To understand automaton A, which is the build mechanism, it is best to start with the Builder as shown in Fig.19.  The active components of the Builder are five Mover blocks expanding forward, three Gluer blocks pointing upward, one Mover block expanding upward, and one Gluer block pointing backward.  The input of the Builder is a stream of ordered blocks coming from the left, as shown in Fig.19(a).  We assume that they are pushed into the Builder by a move-by-five Conveyor (Fig.19(b)).  After the blocks have entered the Builder, the five Movers push these blocks forward (Fig.19(c)).  These blocks then enter the region of influence of the three Gluer blocks (Fig.19(d)).  This glues the blocks to each other and to blocks that had previously entered the Builder, forming a sheet of blocks (Fig.19(e)).  This process continues until the sheet reaches the end of the Builder (Figs.19(e)-(g)).  Once they reach the end, the upward Mover pushes the entire sheet upward (Fig.19(h)).  The sheet enters the region of influence of the single perpendicular Gluer block, which causes it to be glued to sheets that were above it (Fig.19(i)).

\begin{figure}[htbp]
\centering
\includegraphics[scale=0.175]{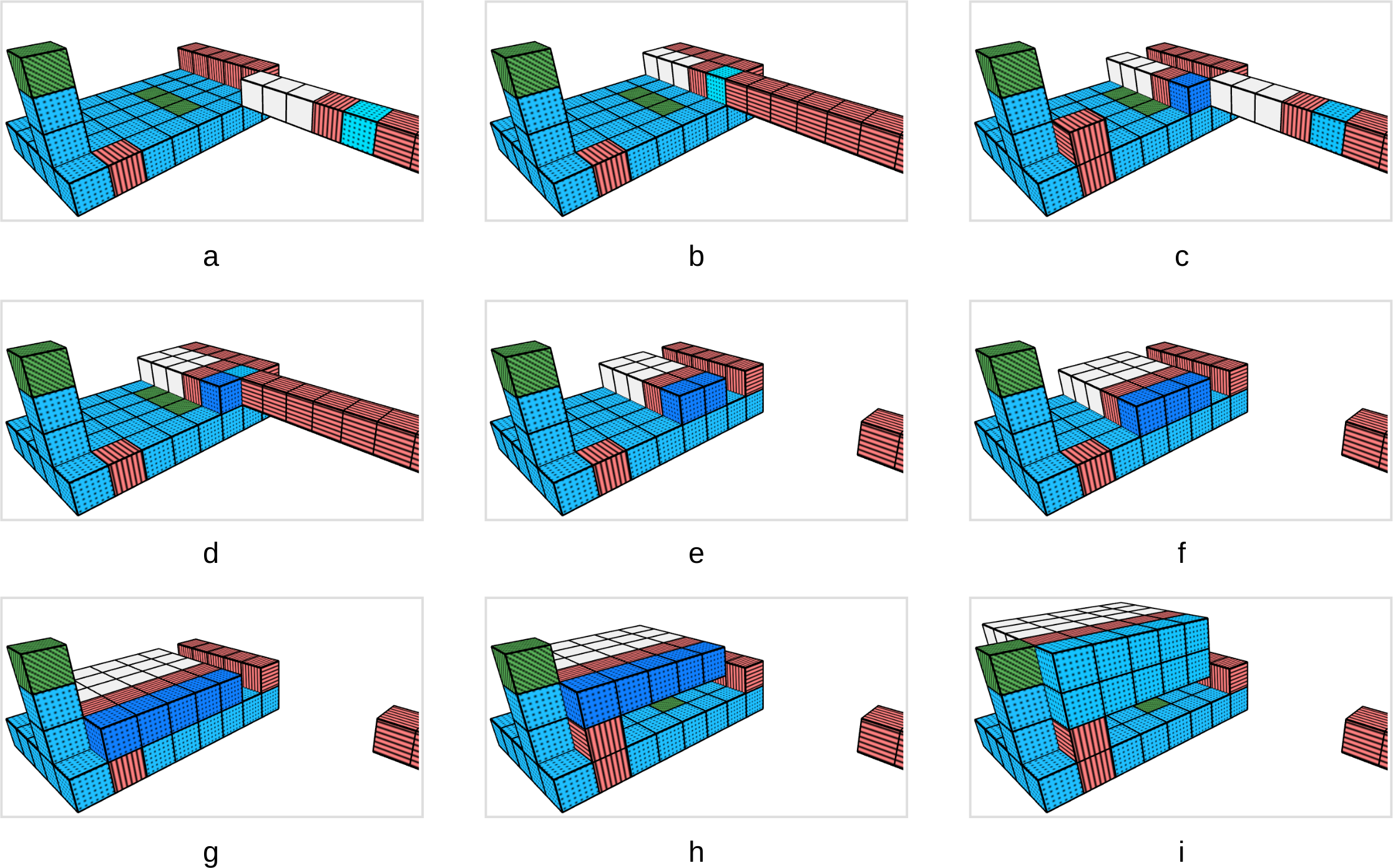}
\caption{Builder}
\label{fig:Fig_19}
\end{figure}

Figure 20(a) shows the Builder building itself.  Notice that, for clarity, we did not show the Dissolvable blocks.  Figure 20(b) shows the Builder building two kinds of 2-tRNA*, with some Dissolvable blocks indicated, but not all Dissolvable blocks shown.

\begin{figure}[htbp]
\centering
\includegraphics[scale=0.175]{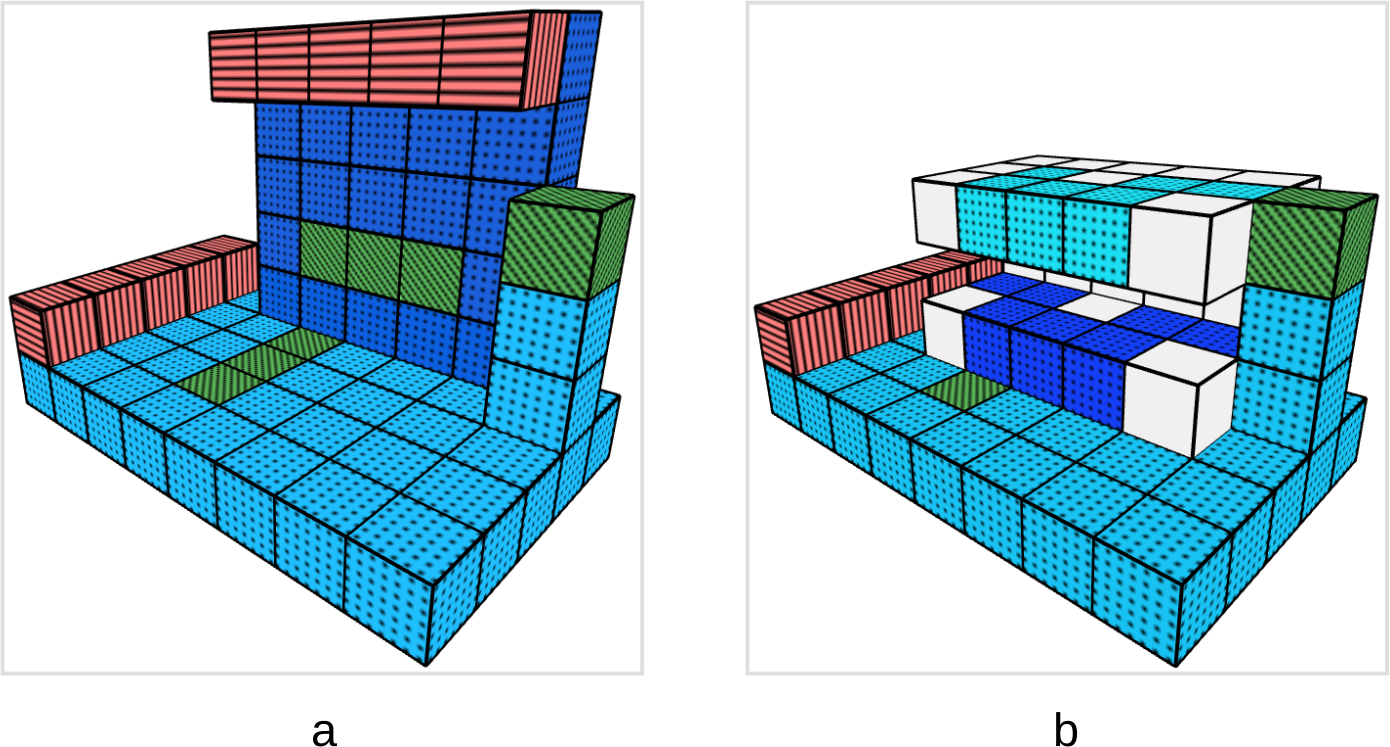}
\caption{Builder building itself and 2-tRNA*}
\label{fig:Fig_20}
\end{figure}

There are a few points to note regarding the Builder presented here.  The Builder can build any three-dimensional structure from the basic building blocks, using Dissolvable blocks as support material to maintain positional information.  The build volume is limited by the number of parallel Movers (five for the Builder in Fig.19) and the length of the building area (five for the Builder in Fig.19).  Therefore, this Builder machine can build any machine with a volume of 5x5xn, where n is arbitrary.  Moreover, the maximum height of the Builder itself is four, given by the position of the perpendicular Gluer block.  

An interesting modification of the Builder is one where the building area has length six.  This Builder can build any machine of volume 5x6xn.  This means that it can build a larger version of itself, for instance, with six parallel Movers and a building area length of six or more.  Since this Builder can build itself and even larger versions of itself, it can be considered universal.

The Builder presented here is not the only type of Builder machine that can be constructed.  For instance, it does not address the issue of removing the finished pieces.  This can be easily accomplished with a Blocker-type mechanism, as shown in the 1-to-2 Converter.  A seemingly serious issue is rotation: the Builder cannot perform rotations; therefore, ideally, all machines and tRNA created are oriented in the same direction.  However, this becomes an issue, especially when the Builder is constructing itself.  If we had many versions of the different machines presented here, with all of them available in different relative 90° rotations with respect to each other, this would not be a serious issue.  In a follow-up publication, we will address this issue in detail.

\subsection{Decoder}

A stream of ordered blocks is required for the Builder to perform its work.  This is what the Decoder machine, as shown in Fig.21, provides.  The input of the Decoder is a tape of RNA (transparent), which enters from the left.  In addition, random 2-tRNA enters the machine from above (Fig.21(a)).  The 2-tRNA is pushed down by something like a move-by-five Conveyor (Fig.21(b)), and then matched against the RNA tape from behind (Fig.21(c)).  Now, two things can occur:

a) There is a match: If it matches, it latches onto the tape, and the move-by-two Conveyor in front moves both the 2-tRNA and the tape (Fig.21(d)).

b) There is no match: If it does not match, the move-by-two Conveyor does nothing, the tape does not move, and the 2-tRNA stays where it was.  In the next pass, this 2-tRNA will be pushed out of the machine downward.

Assuming there was a match, Fig.21(e) depicts what happens next: the left-most Mover pushes the payload off the 2-tRNA, similar to the Redirect pattern discussed earlier.  This creates a stream of blocks whose order depends solely on the encoding of the RNA tape.  After this step, we obtain a stream of ordered blocks, as shown in Fig.21(f), which can be fed into the Builder to construct machines.

\begin{figure}[htbp]
\centering
\includegraphics[scale=0.175]{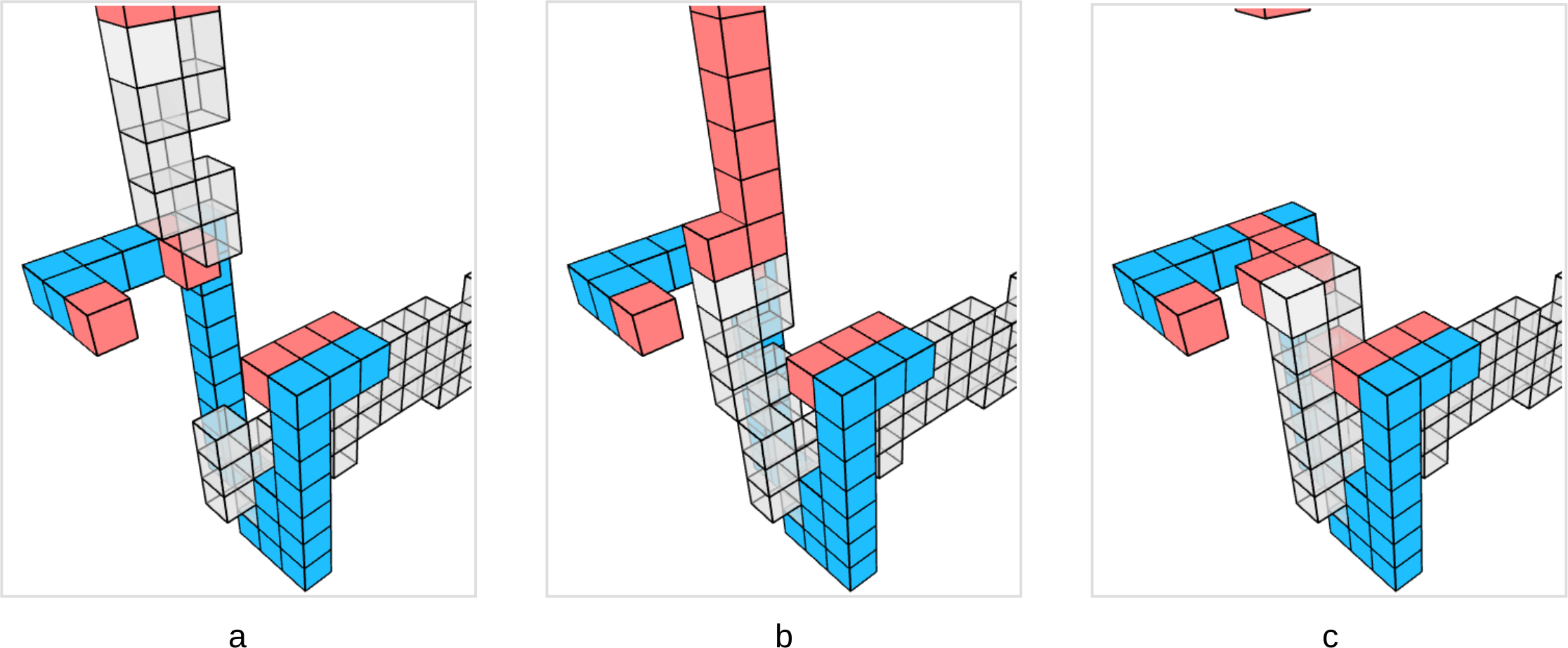}
\includegraphics[scale=0.175]{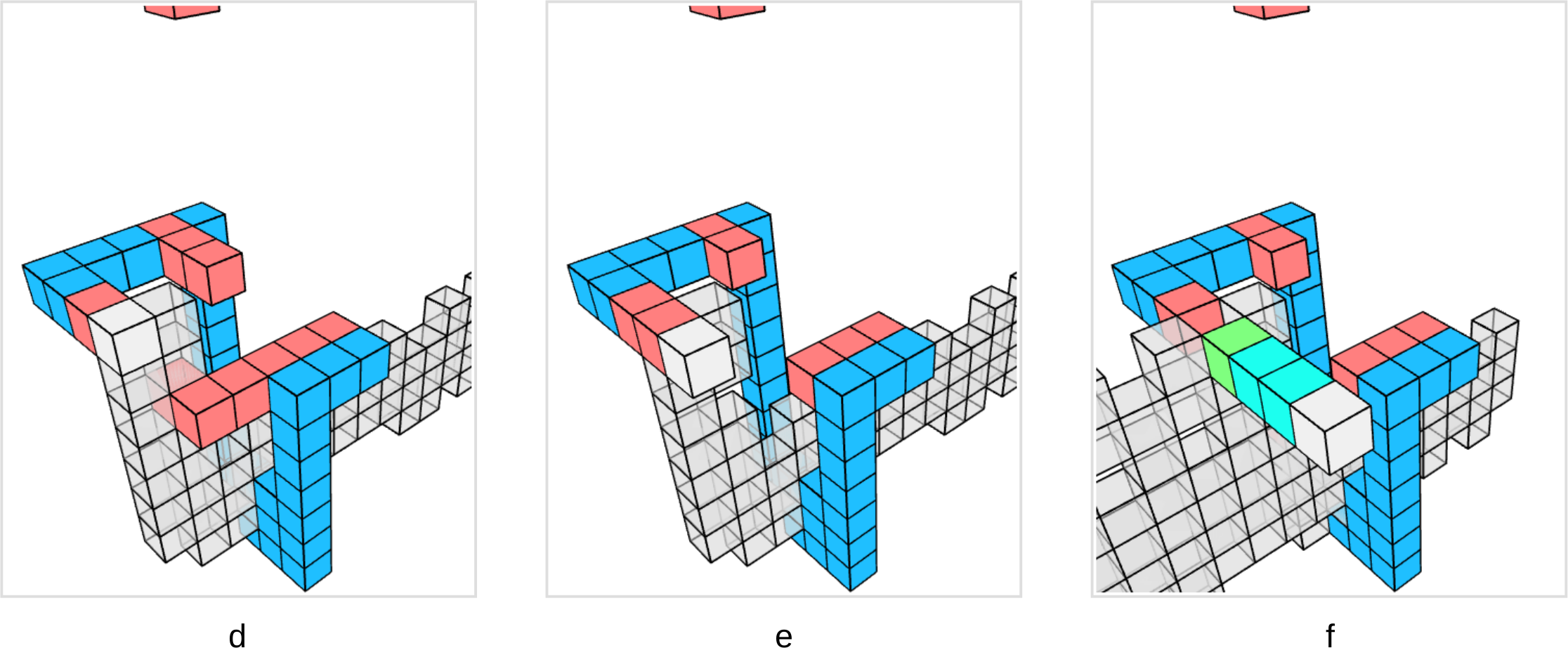}
\caption{Decoder}
\label{fig:Fig_21}
\end{figure}

For this version of the Decoder to work, we have two restrictions: first, we can only use 2-tRNA from Fig.8(b) and Fig.8(c) because they are unique and a unique match is guaranteed.  Second, the tape should start with the codon corresponding to Fig.7(c), so that the Conveyor can grab onto it and move the tRNA together with the tape.  The first restriction means that 2-codons can encode two different blocks, 4-codons four different blocks, 6-codons eight different blocks, and (2n)-codons can encode $2^n$ different block types.

Alternatively, if one is not willing to take the "encoding penalty," one can still use the generic design of the Decoder.  However, in this case, the tRNA entering the Decoder cannot be random but must be sorted in such a way that no confusion can occur.  In addition, the movement of the tape must be synchronized with the length of the codons used.  For instance, if one were to use 4-codons, there are 16 possibilities.  Hence, one must try all 16 possibilities, and only after trying all 16 is one allowed to move the tape forward by one.  Clearly, one needs a counter modulo 16 for this, which, using something like the Back-and-Forth pattern, is doable but non-trivial.

\subsection{Recycler}

The filled 2-tRNA enters the Decoder, and the spent 2-tRNA* leaves the Decoder.  The job of the Recycler, shown in Fig.22, is to refill the spent 2-tRNA* so that they can be used by the Decoder again.  We only sketched the workings of the Recycler.  Figure 22(a) depicts the Recycler for 2-tRNA* (Fig.8(a)) which carries Dissolvable blocks.  Empty 2-tRNA* passes the Recycler on the back, and a matcher (not shown) tries to push them from the back into the Recycler.  Only if 2-tRNA* matches can it be pushed through.  Next, notice the stream of Dissolvable blocks entering the Recycler (Fig.22(a)) from above.  The Mover block pushes one Dissolvable block into the empty space of 2-tRNA*.  To prevent the Mover from continuously pushing the Dissolvable blocks, a blocker mechanism similar to that discussed for the 1-to-2 Converter is needed.  There is one Recycler for each 2-tRNA, as shown in Figs.22(a)-(d), and it is important to match the 2-tRNA* in the correct order.  Assuming that one has 2-tRNA* coming in random order, one should first match for Fig.8(d), then for Fig.8(c) and (b), and finally for Fig.8(a).

\begin{figure}[htbp]
\centering
\includegraphics[scale=0.175]{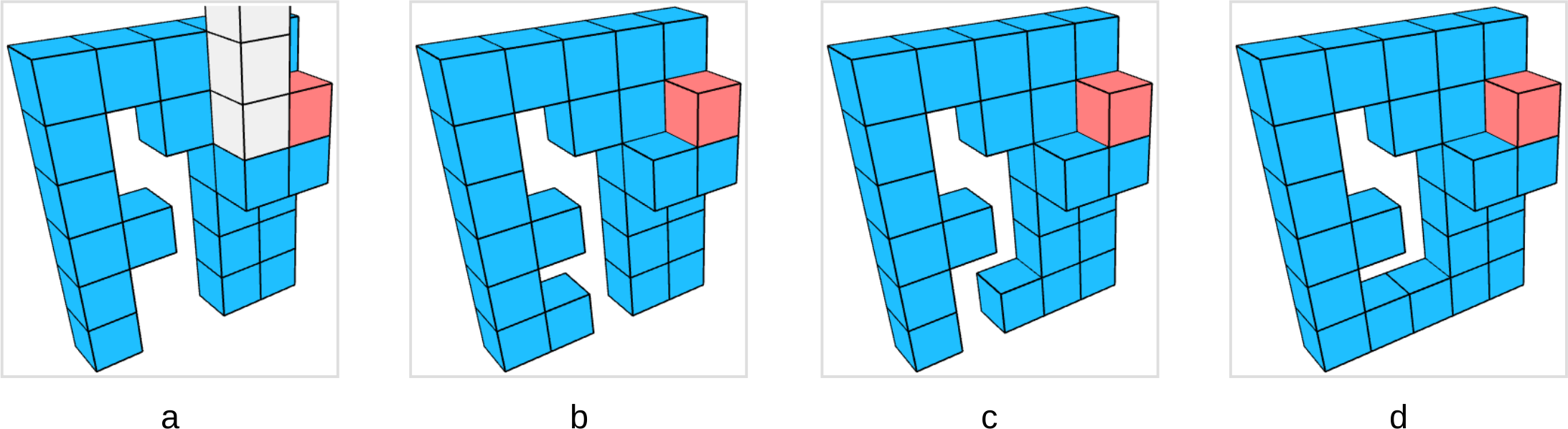}
\caption{Recycler from front}
\label{fig:Fig_22}
\end{figure}

\subsection{Sorter}

The Recycler needs sorted blocks as its input, and it is the Sorter's job to provide them.  The concept behind the Sorter is similar to that of the Matcher pattern introduced earlier.  Figure 23 shows a detailed view of a single Sorter that sorts block types according to their first bit.

Recall that each block has a unique marking through grooves at the bottom.  We see the Sorter block 1000 in the front (yellow), and to the left behind, we see a block of type 0100, which is to be sorted (magenta).  Behind it is a move-by-two (dark blue) and a move-by-one (blue).  In the first step, the move-by-two attempts to push the block over the Sorter block.  This is only successful if the first bit of the block to be sorted is zero; otherwise, it is blocked.  If blocked, the move-by-one moved it out of the way.  In the next step, if we match against the Sorter block 0100, we can test for the second bit.  In the final step, we test for the third bit.

\begin{figure}[htbp]
\centering
\includegraphics[scale=0.175]{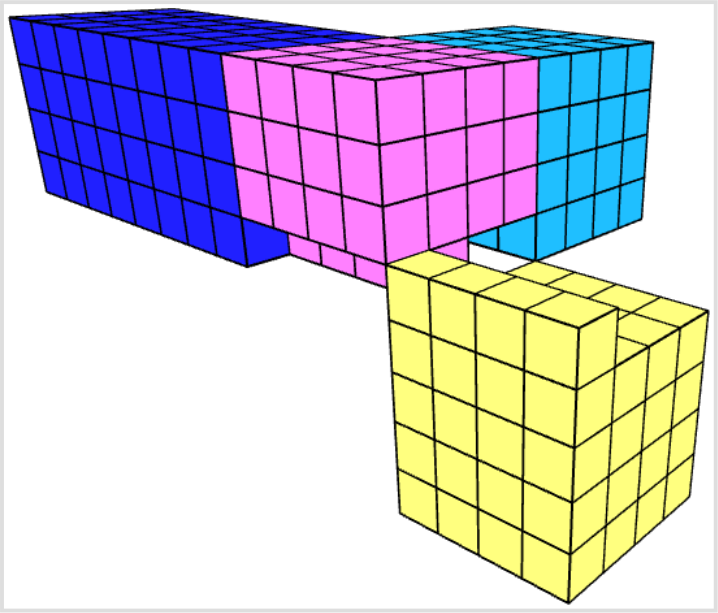}
\caption{Single Sorter}
\label{fig:Fig_23}
\end{figure}

After matching, the block will be in either one of the two transparent locations, as depicted in the zoomed-out view (Fig.24(a)).  If the first bit is zero, the block will have moved to the transparent location on the right; otherwise, it will move to the transparent location on the left.  The blocks enter the Sorter from above and are pushed down by another move-by-two from above.

\begin{figure}[htbp]
\centering
\includegraphics[scale=0.175]{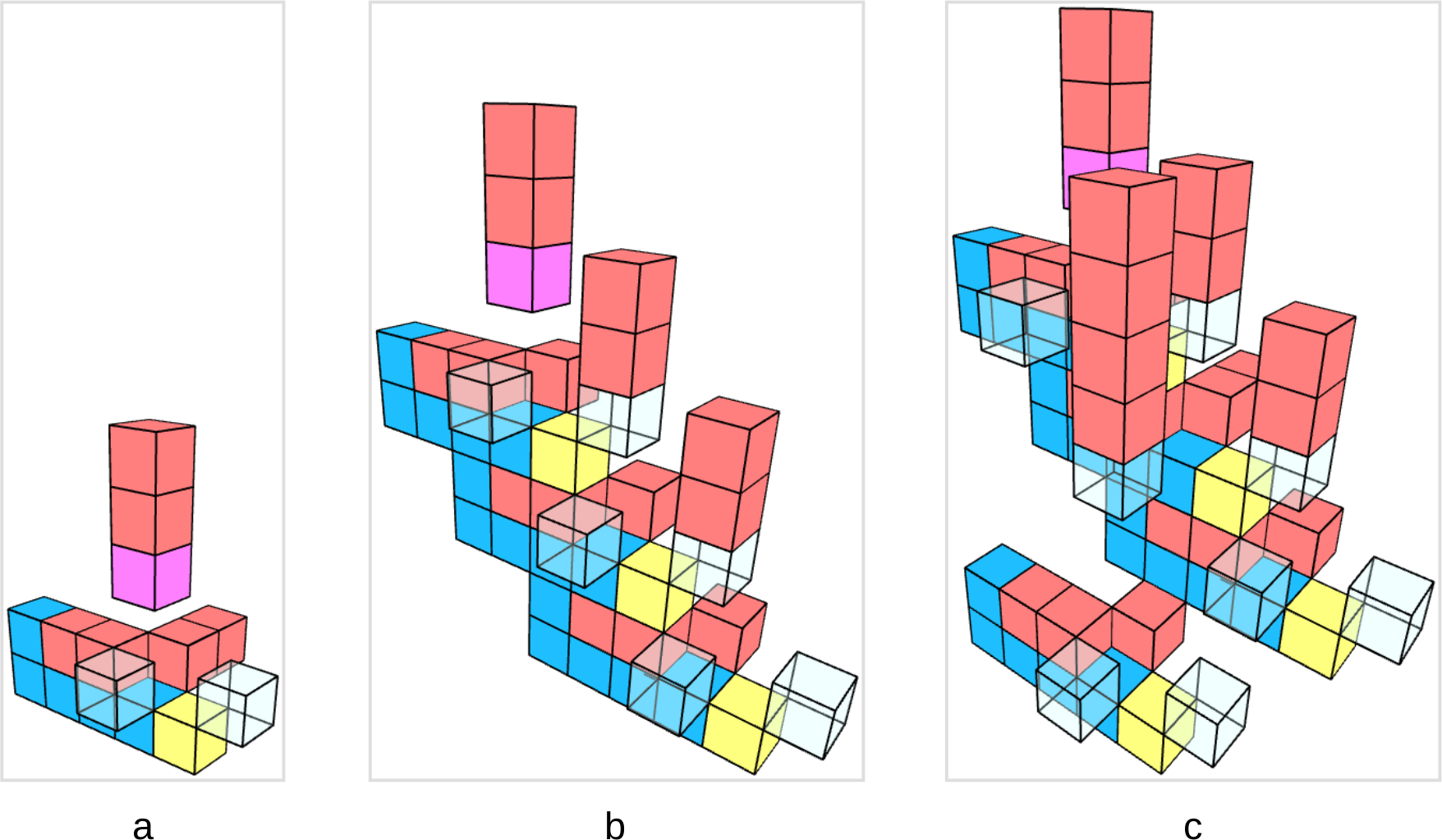}
\caption{Sorter machines}
\label{fig:Fig_24}
\end{figure}

Figure 24(b) depicts three of these simple Sorters in a row, each with a different Sorter block, thus testing for bit zero, bit one, and bit two.  Finally, we find Sorter block type 0001 at the rightmost transparent location and block type 0010 at the next rightmost transparent location.

Figure 24(c) shows that by adding another Sorter in front, we can identify block types 0100 and 0110 at the two lowermost transparent locations.  By adding three more sorters, we can distinguish eight block types.

We have thus indicated how automaton A, von Neumann's universal constructor, can be built from the Sorter, Recycler, Decoder, and Builder machines.

\subsection{Ambiguities}

Note on the efficiency of the Sorter machine: We need to know the exact three-dimensional orientation of the blocks.  Hence, the grooves can only be on one of the six sides of the cube.  Because the blocks might be randomly oriented, only one in six will be oriented such that the grooves are at the bottom.  Additionally, the blocks can be rotated around the z-axis in four different ways, only one of which is correct.  Hence, only 1/6 * 1/4, that is about 4\%, of the blocks will be recognized.  The efficiency can be significantly increased for blocks with no orientation, such as Dissolvable and Simple blocks.  For these, all six sides could have grooves, and being symmetric, the efficiency would be 50\%.

An interesting question is which encoding scheme should be choosen.  Naively, one would want to allow for all possible combinations.  If we were to do that, then for the four Sorter blocks depicted in Fig.5, we would have $2^4$, or 16 possibilities. However, two factors must be considered.

Because we do not know in which orientation blocks enter the Sorter, we need to exclude the "1111" encoding, as we could not distinguish it from a block that has entered the Sorter with the wrong orientation.  We can also exclude the "0000" encoding because the blocks would only be 75\% of the standard height, causing various problems.

Assuming a block enters the Sorter with the correct orientation, we still have one problem: except for the Dissolvable and Simple blocks, all other blocks have directional information. Table 1 lists each combination with its respective symmetric partner, which is a block rotated by 180° around the z-axis.

\begin{table}[htbp]\small
\caption{Symmetries}
\label{tab:Tab_01}
\begin{center}
\begin{tabular}{|c|}
\hline
0000 = 0000\\
0001 = 1000\\
0010 = 0100\\
0011 = 1100\\
0101 = 1010\\
0110 = 0110\\
0111 = 1110\\
1001 = 1001\\
1011 = 1101\\
1111 = 1111\\\hline
\end{tabular}
\end{center}
\end{table}

Having already excluded the first and last choices, we must select only one per row from the remaining eight in Table 1.  The specific choice does not matter.  As shown in Fig.6, we selected the following encoding:

\begin{center}
\begin{minipage}{0.8\textwidth}
\begin{verbatim}
0001, 0011, 0100, 0110, 1001, 1010, 1101, 1110
\end{verbatim}
\end{minipage}
\end{center}

Note that 0110 and 1001 are symmetric in themselves (palindromes), meaning they cannot carry directional information.  However, if we use them to encode the Dissolvable and Simple blocks, they can be used for encoding.

Hence, using this schema, we can uniquely distinguish eight different block types with four Sorter blocks.  As indicated above, this is not sufficient; we must be able to distinguish at least 16 different block types.  Therefore, we need to generalize our encoding scheme to a higher number of bits.  Careful counting considerations lead to Eq.(\ref{eq:eq01})
for the number of distinguishable blocks

\begin{equation}
f(n) = 
\begin{cases} 
    2^{n-1} + 2^{(n-1)/2} & \text{if } n \text{ is odd}, \\
    \frac{1}{2} \left( 2^n + 2^{n/2} \right) & \text{if } n \text{ is even}.
\end{cases}
\label{eq:eq01}
\end{equation}

If we wanted to remove palindromes, we would have to subtract $2^{n/2}$ from the above for even n, and twice that for odd n, that is, $2^{(n+1)/2}$.

\begin{table}[htbp]\small
\caption{Number of blocks}
\label{tab:Tab_02}
\begin{center}
\begin{tabular}{|c|c|c|}
\hline
$n$ & $f(n)$ & $g(n)$\\\hline
 2  &    3  &    1\\
 3  &    6  &    2\\
 4  &   10  &    6\\
 5  &   20  &   12\\
 6  &   36  &   28\\
 7  &   72  &   56\\
 8  &  136  &  120\\\hline
\end{tabular}
\end{center}
\end{table}

Table 2 shows the number of distinguishable blocks depending on the number of Sorter blocks n.  The f(n) column lists the total number allowing for palindromes, whereas the g(n) column lists the total number if we do not allow for palindromes.  Because we must be able to distinguish around sixteen different block types, we need at least five different Sorter blocks if we can use palindromes, and six if we cannot.

\subsection{Jacquard Machine}

Several more machines can be built using the proposed mechanism.  They are classified as automata D in von Neumann's language.  In homage to Joseph Marie Jacquard, we show a simple implementation of a Jacquard machine based on our Back-and-Forth pattern.  Figure 25 shows a slightly modified version of the same principle that creates different types of repetitive patterns.  The generalization to longer periods and larger widths is straightforward.  These can be used to make sticks, codons, patterns, tracks, or walls.

\begin{figure}[htbp]
\centering
\includegraphics[scale=0.175]{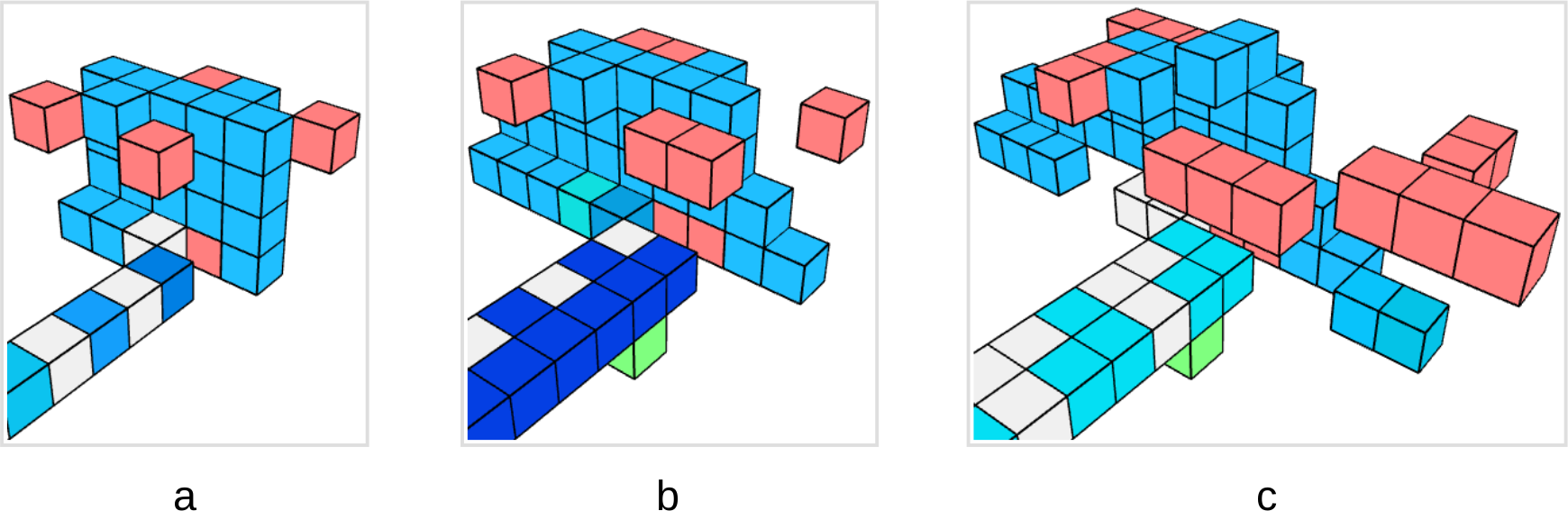}
\caption{Jacquard machines}
\label{fig:Fig_25}
\end{figure}

Figure 25(a) shows the simplest version, creating a pattern of period two and width one.  Figure 25(b) shows a Jacquard machine that creates a pattern of period two and width two.  The machine in Fig.25(c) creates a pattern of period three and width two.  Period three is generated by having two move-by-twos that move to the left, and one move-by-four that moves to the right.  It is straightforward to generalize this to longer periods and larger widths.

\section{Discussion}

We have shown how to construct a self-replicating mechanical system using as few assumptions as possible.  Essentially, this is a mechanical replication of what occurs inside living biological cells.  Our approach is inspired by the physical world; however, it is still a theoretical model with some severe restrictions, formulated in rules 1-14.  It is inspired by biology and is supported by computer simulations.  The aim of this work is not to demonstrate real working physical machines, but to show that the problem of self-replication is tractable and within the reach of current technology.

Even without the context of self-replication, the machines and mechanisms presented here are of individual interest.  Anyone familiar with the idea of molecular manufacturing [6,7], will recognize the potential of Jacquard machines in terms of smart materials.  The Builder, being a programmable construction machine, can also be considered in this context.

We also briefly mentioned that our architecture allows for convergent assembly, that is, building machines at a larger scale mimicking themselves, as suggested in [6-8].  The mechanisms suggested in references [10] and [11] mainly focus on the assembly.  The basic building blocks in both publications have significantly higher capabilities and complexity than our proposed basic building blocks.  Neither addresses the issue of sorting the basic building blocks, and most importantly, in both references, the assembly process was controlled through software and external processors.

Returning to von Neumann [1]: We attempted to decompose the problem of self-replication into its core components, particularly automaton A and automaton B.  Additionally, we further decomposed automaton A and automaton B into small machines, each performing a single task, as indicated in Fig.1.  For each of these small machines, we described their workings and composition, showing how they can be built from a set of five different block types, each having one particular property.

We also show how information can be stored in the form of a tape of codons, which is von Neumann's description function $\Phi$.  We described how that information can be replicated and translated into building instructions via tRNA.

We were not trying to minimize the number of block types, but from the detailed discussion above, it is clear that there must be some basic building functionality, which entails Simple and Glue blocks, some movement functionality, and some sorting functionality.  Sorting functionality entails several block types, such as movement functionality owing to the timing and directional requirements.  Hence, it is impossible to reduce the number of required block types to less than eight, and only with some imaginative tricks below 16. 

However, the number of block types directly drives the number of machines, because both the number of Recyclers needed and the size of the Sorter are directly proportional to the number of block types.  Most of our machines fit into a volume of approximately 4x5x6 blocks, or approximately 120 blocks.  Assuming eight different block types, we needed seven Sorters, eight Recyclers and one each of Stick-Maker, Copier, Decoder and Builder, totaling 19 machines.  Thus, a minimum of approximately 120*19, that is, 2280 blocks are required for building automata A and B.  Assuming that we can use 4-codons to encode the building instructions, and assuming each 4-codon consists of eight blocks, the RNA tape would consist of 72960 blocks.  In general, the number of blocks for building the machines depends on the number n of block types such as $O(n)$, and the size of the genome grows like $O(n \log n)$.

Also interesting is the dependence of the timing on the number of block types.  The number of block types has no influence on the timing behavior of the Builder.  For the Sorter, there are two aspects: one is due to geometry, where, because we have the markings only on one of the six sides and the orientation is important, only 4\% of blocks are recognized in one run.  Hence, on an average, 25 comparisons must be made to recognize a particular block type.  However, this number does not depend on the number of block types.  The second aspect is the sorting itself.  Because the sorting mechanism is based on a tree-type structure, the time needed to pass through the Sorter will only grow logarithmically with the number of block types.  In addition, because the encoding scheme grows exponentially, it has little effect on timing, but it should have a linear effect on the error rate.  The timing of the Recycler is linear with the number of block types.  As for the Decoder, the timing behavior is also linear in block types, because the sorting has effectively been done by the Recycler owing to the nature of its arrangement.  The real problem is the combination of the Recycler and Decoder: unless one is willing to pay the encoding penalty, the timing behavior of the combination of the Recycler and Decoder will grow quadratically with the number of block types, that is, $O(n^2)$. 

No mechanical system is perfect, and errors occur in the real world. An interesting question is how errors and mutations affect the proposed mechanism and how fault-tolerant it is. One of the biggest issues is jamming, where the functioning of one machine is prevented, and because many of the machines depend on each other, the entire mechanism will stop working. Parallelization could be a solution here, although due to the intricate dependency of the Sorter, Recycler, and Decoder, the proposed mechanism has only limited potential for parallelization.  Mutations are another source of errors. The functionality of the presented machines depends on the location and orientation of the active components, that is, the Movers, the Gluers, and the Sorters. All other changes have little influence. Hence, single-bit flips should have a minor effect. However, deletions or additions of bits in RNA have profound effects. Another effect should arise from wear and tear, that is, degradation over time. This depends heavily on the concrete implementation of the mechanism; hence, no estimates can be provided.

For many of the machines discussed here, we only show the core mechanism and elaborate only on the required support structure.  In von Neumann's language, this is automaton C, which is clearly non-trivial and deserves further investigation.  We only show how to create the machines and how each machine works individually.  In Fig.1 we show that the output of one is the input of the other.  However, we do not show how to arrange the machines in a proper manner such that they work smoothly together.  We do not show how the components move from one machine to another.  We do not show how they obtain their respective relative positions and how they get assembled.  We only indicate the mechanisms for controlling, starting, and stopping the machines.  It appears that automaton C is significantly more complex than von Neumann indicated.

\section{Conclusion}

Despite being very simplistic, sketching and skipping many implementation details, the mechanism shown here can help to elucidate and answer a few questions of interest regarding self-replicating systems.  It also provides interesting constraints that do not depend on any particular technology.  There remain many aspects worth investigating, especially with respect to automaton C.  We have made several suggestions related to automaton C, but further investigation is required.  The transport of material, assembly, and relative positioning of the machines, as well as the control of the machines, are important issues.  The building mechanism itself seems to be rather wasteful, and a solution to the orientation issue of the machines needs to be determined.  Another interesting aspect might be the disassembly and reuse of broken machines and materials.

Finally, it is interesting to point out that our mechanism essentially performs the operations of sorting, copying, reading, and, to a limited extent, writing; that is, it performs information processing.  This is worthy of further investigation.  According to Sydney Brenner [12], biologists ask only three questions of a living organism: How does it work?  How is it built?  How did it get that way?  In our model, we answer the first two questions.  The third question should be addressed in future research.

\section{Acknowledgement}
I am thankful to L. Ochs and W.Z. Taylor for discussions, feedback, and suggestions.

%\section{Appendix}
\appendix
\section{Appendix}

In the following, we describe all machines in terms of their MDL.

\subsection{Fig.10:}
\begin{verbatim}
// z=0:
M11
M11
M30
\end{verbatim}

\subsection{Fig.11:}
\begin{verbatim}
// z=0:
M02
b__
M11
M11
M00
\end{verbatim}

\subsection{Fig.13:}
\begin{verbatim}
// z=0:
      b__b__b__
M26b__b__

// z=1:
            M38


// z=2:
            b__


// z=3:
M52b__b__b__b__
\end{verbatim}

\subsection{Fig.14:}
\begin{verbatim}
// z=0:
b__b__b__b__

// z=1:
b__      b__

// z=2:
b__

// z=3:
M00
\end{verbatim}

\subsection{Fig.15:}
\begin{verbatim}
// z=0:
   b__
b__b__
   b__
   b__b__
   b__
\end{verbatim}

\subsection{Fig.16:}
\begin{verbatim}
// z=0:
b__b__b__b__b__b__b__b__b__b__G2_

// z=1:
b__M02M02M02M02
\end{verbatim}

\subsection{Fig.18:}
\begin{verbatim}
// z=0:
b__b__b__b__b__b__b__
                  b__

// z=1:

                  b__


// z=2:

                  b__

// z=3:

                  b__

// z=4:

                  b__

// z=5:

                  b__

// z=6:
b__M04M04
                  G4_

// z=7:
b__
                  b__

// z=8:
b__b__b__M51M51b__
               b__b__
\end{verbatim}

\subsection{Fig.19:}
\begin{verbatim}
// z=0:
b__b__b__b__b__b__M29b__
b__b__b__G2_b__b__b__b__
b__b__b__G2_b__b__b__b__
b__b__b__G2_b__b__b__b__
b__b__b__b__b__b__b__b__

// z=1:
M02
M03                  b__
M04
M05
M06

// z=2:

                     b__

// z=3:

                     G3_
\end{verbatim}

\subsection{Fig.21:}
\begin{verbatim}
// z=0:
      b__
      b__
   b__b__
   b__
   b__
   b__

// z=1:
      b__
      b__



   b__

// z=2:
      b__
      b__



   b__

// z=3:
      b__




   b__

// z=4:
      b__




   b__

// z=5:
      b__




   b__

// z=6:
b__b__b__
M02M02M11



   b__

// z=7:





   b__

// z=8:



               M44
      M00      b__
   b__M41b__b__b__
\end{verbatim}

\subsection{Fig.22a:}
\begin{verbatim}
// z=0:

b__b__      b__

// z=1:

b__b__      b__

// z=2:

b__b__   b__b__

// z=3:
b__b__
b__b__      b__

// z=4:
M00
b__b__b__   b__

// z=5:

b__b__b__b__b__
\end{verbatim}

\subsection{Fig.24b:}
\begin{verbatim}
// z=0:
S__b__b__b__b__


// z=1:
      M33M33b__
   M47

// z=2:
      S__b__b__b__b__


// z=3:
            M33M33b__
         M47

// z=4:
   M50      S__b__b__b__b__


// z=5:
   M50            M33M33b__
               M47

// z=6:
         M50

// z=7:
         M50

// z=8:
               M50

// z=9:
               M50
\end{verbatim}

\subsection{Fig.25:}
\begin{verbatim}
// z=0:
   b__
   b__




      b__
      b__
   b__b__

// z=1:

   b__
   b__
   b__
   b__
   b__
   b__
   b__
   b__

// z=2:


   b__
   b__
   b__
   b__
   b__

// z=3:


   b__
b__b__
   b__
   b__b__
   b__
\end{verbatim}

\nocite{*}
\bibliographystyle{fundam}
\bibliography{citations}

\begin{thebibliography}{10}
\providecommand{\url}[1]{\texttt{#1}}
\providecommand{\urlprefix}{URL }
\expandafter\ifx\csname urlstyle\endcsname\relax
  \providecommand{\doi}[1]{doi:\discretionary{}{}{}#1}\else
  \providecommand{\doi}{doi:\discretionary{}{}{}\begingroup
  \urlstyle{rm}\Url}\fi
\providecommand{\eprint}[2][]{\url{#2}}

\bibitem{p1r01}
Von~Neumann J, Burks AW, et~al.
\newblock Theory of self-reproducing automata.
\newblock 1966.

\bibitem{p1r02}
Alberts B, Heald R, Johnson A, Morgan D, Raff M, Roberts K, Walter P.
\newblock Molecular Biology of the Cell: Seventh International Student Edition
  with Registration Card.
\newblock WW Norton \& Company, 2022.

\bibitem{p1r03}
Nelson DL, Lehninger AL, Cox MM.
\newblock Lehninger principles of biochemistry.
\newblock Macmillan, 2008.

\bibitem{p1r04}
O’Donnell M, Langston L, Stillman B.
\newblock Principles and concepts of DNA replication in bacteria, archaea, and
  eukarya.
\newblock \emph{Cold Spring Harbor perspectives in biology}, 2013.
\newblock \textbf{5}(7).

\bibitem{p1r05}
Kristoffersen EL, Burman M, Noy A, Holliger P.
\newblock Rolling circle RNA synthesis catalyzed by RNA.
\newblock \emph{Elife}, 2022.
\newblock \textbf{11}.

\bibitem{p1r06}
Drexler KE.
\newblock Nanosystems: molecular machinery, manufacturing, and computation.
\newblock John Wiley \& Sons, Inc., 1992.

\bibitem{p1r07}
Freitas RA, Merkle RC.
\newblock Kinematic self-replicating machines.
\newblock Landes, 2004.

\bibitem{p1r08}
Merkle RC.
\newblock Convergent assembly.
\newblock \emph{Nanotechnology}, 1997.
\newblock \textbf{8}(1):18.

\bibitem{p1r09}
Penrose LS.
\newblock Self-reproducing machines.
\newblock \emph{Scientific American}, 1959.
\newblock \textbf{200}(6):105--117.

\bibitem{p1r10}
Collins CM.
\newblock Self reproducing fundamental fabricating machine system, 1998.
\newblock US Patent 5,764,518.

\bibitem{p1r11}
Langford WK, Ghassaei A, Gershenfeld N.
\newblock Self-assembling assemblers and manipulators built from a set of
  primitive blocks, 2018.
\newblock US Patent 10,155,314.

\bibitem{p1r12}
Brenner S.
\newblock Life's code script.
\newblock \emph{Nature}, 2012.
\newblock \textbf{482}(7386):461--461.

\end{thebibliography}

%%%%%%%%%%%%%%%%%%%%%%%%%%%%%%%%%%%%%%%%%%%%%%%%%%%%%%%%%%%%%%%%%%%%%%

\end{document}